\documentclass[11pt,a4paper]{article}
\pdfoutput=1
\usepackage{jcappub}
\usepackage[utf8]{inputenc}
\usepackage{aas_macros}
\usepackage{amsfonts}
\usepackage{amsmath}
\usepackage{bm}
\usepackage{amssymb}
\usepackage{graphicx}
\usepackage[dvipsnames]{xcolor}
\usepackage{xfrac}
\usepackage{hyperref}
\usepackage{nameref}
\usepackage{cleveref}
\usepackage{multirow}
\usepackage{soul}
\usepackage{url}
\usepackage[thinlines]{easytable}
\crefformat{section}{\S#2#1#3}
\hypersetup{colorlinks=true,allcolors=teal}

\defcitealias{Arya_2023}{A23}
\defcitealias{Arya_2024}{A24}

\providecommand{\keywords}[1]
{
  \small	
  \textbf{\textit{Keywords---}} #1
}

\newcommand{\Mpch}{\ensuremath{h^{-1}{\rm Mpc}}}
\newcommand{\kpch}{\ensuremath{h^{-1}{\rm kpc}}}

\newcommand{\be}{\begin{equation}}
\newcommand{\ee}{\end{equation}}

\title{\boldmath A trailing lognormal approximation of the Lyman-$\alpha$ forest: comparison with full hydrodynamic simulations at $2.2\leq z\leq 2.7$}

\author[]{B. Arya}

\affiliation[]{Indian Institute of Technology, Kanpur 208016, Uttar Pradesh, India}

\emailAdd{bhaskara@iitk.ac.in}

\abstract{Lyman-$\alpha$(Ly$\alpha$) forest in the spectra of distant quasars encodes the information of the underlying cosmic density field at smallest scales. The modelling of the upcoming large and high-fidelity forest data using cosmological hydrodynamical simulations is computationally challenging and therefore, requires accurate semi-analytical techniques. One such approach is based on the assumption that baryonic density fields in the intergalactic medium (IGM) follow lognormal distribution.
Keeping this in mind, we extend our earlier work to improve the lognormal model of the Ly$\alpha$ forest in recovering the parameters characterizing IGM state, particularly the hydrogen photoionization rate ($\Gamma_{12}$), between $2.2 \leq z \leq 2.7$, by simulating the model spectra at a slightly lower redshift than the Sherwood smooth particle hydrodynamical simulations (SPH) data. The recovery of thermal parameters, namely, the mean-density IGM temperature ($T_0$) and the slope of the temperature-density relation ($\gamma$) is also alleviated. These parameters are estimated through a Markov Chain Monte Carlo (MCMC) technique, using the mean and power spectrum of the transmitted flux.
We find that the usual lognormal distribution of IGM densities tend to over-predict the number of Ly$\alpha$ absorbers seen in SPH simulation. A lognormal model simulated at a lower redshift than SPH data can address this limitation to a certain extent. We show that with such a "trailing" model of lognormal distribution, values of $\Gamma_{12}$ are recovered at $\lesssim 1-\sigma$. We argue that this model can be useful for constraining cosmological parameters.}


\begin{document}
\label{firstpage}
\maketitle
\flushbottom

\section{Introduction}
\label{sec:intro}

The Lyman-$\alpha$ forest seen in the spectra of distant quasi-stellar objects (QSOs) are sensitive to the underlying cosmic density field at the smallest scales \citep{Rauch_1998, Weinberg_2003, Meiksin_2009, McQuinn_2016}. The forest provides insight into the thermal and ionization of the intergalactic medium (IGM) \citep{2000bgfp.conf..455S, Bolton_2005, gaikwad2020, Gaikwad_2021}, and is used for constraining cosmological \citep{1992AN....313..265L, 1998APS..APR.8WK05W, 2003MNRAS.342L..79S, 2003MNRAS.344..776M, Bird2023priya, tohfa2023forecast, Khan_2023} and dark matter models \citep{hansen_2002, Viel_2005, baur_2017, irsic_17b, bose_2019, garzilli_2019, palanque_2020, pedersen_2020, garzilli_2021, sarkar_2021}. The need for efficient computational modelling of astrophysical and cosmological parameters has become crucial with the advent of large cosmological surveys such as ongoing DESI \citep{ribera_2018, karacayli_2020, walther_2021, desi_2022, satya_2022} and upcoming WEAVE \citep{dalton_2012, weave_2022}.

One method to accomplish this is to use semi-analytical models that approximate baryonic density field, and use physical parameters to generate Ly$\alpha$ spectra (see \citep[][henceforth, A23, A24 respectively]{Arya_2023, Arya_2024} for such examples). With these methods, it is possible to jointly explore the astrophysical and cosmological parameters. This is particularly important when considering parameters related to dark matter phenomenology (such as, e.g., the mass of a `warm' dark matter candidate) which lead to suppression of power at small scales, since such effects may also arise due to variations in the thermal history of the IGM.

This work is third in the series of papers where we have attempted to exploit the efficiency of the lognormal approximation of the baryonic density field to constrain astrophysical and cosmological parameters \citep{Arya_2023, Arya_2024}. In our earlier works, we integrated the model with an end-to-end MCMC analysis method, which offers a quick and simplistic way of modelling the IGM. We tested the model in recovering the thermal and ionization parameters against Sherwood simulation, a SPH simulation, at $z \sim 2.5$ in \citetalias{Arya_2023}, and then extended the work to other redshifts, $z \epsilon \{2, 2.1, 2.2, 2.3, 2.4, 2.5, 2.6, 2.7\}$ while improving upon methodology and introducing additional parameter, $\nu$, to scale the 1D baryonic density field. In \citetalias{Arya_2024}, we had shown that while the thermal parameters (IGM temperature at mean density, $T_0$ and equation of state index, $\gamma$) are estimated reliably at $\lesssim 1-\sigma$ for most redshifts, the model could not recover photoionization rate ($\Gamma_{12}$), with the discrepancy at $\gtrsim 3-\sigma$ for $z > 2.2$.
Therefore, in this paper, we present a "trailing" lognormal approximation of the density field, where we simulate the model at a lower redshift than the data. We show that such a model shows significant improvement in the recovery of $\Gamma_{12}$ at $z \geq 2.2$ with the best-fit $\Gamma_{12}$ at $\lesssim 1-\sigma$ from the true value. At the same time, the model is also able to improve the recovery of the thermal parameters. This method also does not require parameter $\nu$, which simplifies the setup and speeds up the parameter estimation. We further investigate the reasons behind it and comment on its usefulness in joint astrophysical and cosmological parameter space exploration simultaneously in multiple redshift bins.

The layout of the paper is as follows, in \cref{sec:prep}, we briefly outline the methodology for calculating flux statistics, covariance matrices, performing likelihood analysis as well as describe the trailing lognormal model. In \cref{sec:result}, we present our results of recovering thermal and ionization histories and also discuss the reasons behind the improvement in lognormal model and eventually conclude in \cref{sec:conclude}.

\section{Simulations \& Method}
\label{sec:prep}
\noindent
We use the Ly$\alpha$ forest from the Sherwood SPH simulations as benchmark for validating our model and request the readers to refer to \citetalias{Arya_2023} for more details.
Throughout this work, we fix cosmological parameters for lognormal to Planck 2014 cosmology, the same being used in Sherwood simulations, \{$\Omega_m = 0.308$, $\Omega_{\Lambda} = 1 - \Omega_m$, $\Omega_b = 0.0482$ $h = 0.678$, $\sigma_8 = 0.829$, $n_s = 0.961$, $Y = 0.24$\}, consistent with the constraints from \citep{Planck_2014}. 

\subsection{Lognormal approximation}
In this section, we briefly describe the procedure to generate Ly$\alpha$ spectra using the lognormal model. We obtain the the 3D baryonic power spectrum, $P^{(3)}_{\mathrm{b}}(k, z)$ at any given redshift $z$, by smoothing the linear DM density power spectrum , $P_{\mathrm{DM}}(k, z)$, \footnote{We use the CAMB transfer function \citep[\url{https://camb.readthedocs.io/en/latest/}]{camb} to calculate linear matter power spectrum for a given set of cosmological parameters, same as Sherwood simulations \citep{bolton+17-sherwood}.} over Jeans length $x_{\textrm{J}}(z)$,\footnote{Unlike in some literature \citep{Kulkarni_2015, rorai_2017}, where smoothing is done on the Ly$\alpha$ transmitted flux, we use a more physical way by smoothing the DM density field itself.}
\begin{equation}
    P^{(3)}_{\mathrm{b}}(k, z) = D^2(z) P_{\mathrm{DM}}(k)~\mathrm{e}^{-2 x_{\mathrm{J}}^2(z) k^2}.
    \label{eq:Pb_PDM_gaussian}
\end{equation}
where $D(z)$ is the linear growth factor. The assumption here is that the baryons trace the dark matter at large scales $k^{-1} \gg x_{\mathrm{J}}$ and are smoothed because of pressure forces at scales $k^{-1} \lesssim x_\mathrm{J}$. Since the Ly$\alpha$ forest probes the cosmic fields only along the lines of sight, it is sufficient to generate the line of sight baryonic density field $\delta_{\mathrm{b}}^L(x, z)$ and the corresponding line of sight component of the velocity fields $v_{\mathrm{b}}^L(x, z)$. We can obtain the 1D baryonic $(P^{(1)}_{\mathrm{b}}(k, z))$ and linear velocity $(P^{(1)}_{\mathrm{v}}(k, z))$ power spectra from 3D baryonic power spectra by
\begin{equation}
    P^{(1)}_{\mathrm{b}}(k, z) = \frac{1}{2\pi}\int_{|k|}^{\infty} dk^{\prime} k^{\prime} P^{(3)}_{\mathrm{b}}(k, z),
    \label{eq:Pb_1d}
\end{equation}and 
\begin{equation}
    P^{(1)}_{\mathrm{v}}(k, z) = \dot{a}^2(z) k^2 \frac{1}{2\pi}\int_{|k|}^{\infty} \frac{dk^{\prime}}{k^{\prime 3}} P^{(3)}_{\mathrm{b}}(k, z),
    \label{eq:Pv_1d}
\end{equation}
where $a$ is the scale factor and $\dot{a}$ is given by the Friedman equation
\begin{equation}
    \dot{a}^2(z) = H_0^2\left[\Omega_{\mathrm{m}}(1 + z) + \Omega_{\mathrm{k}} + \frac{\Omega_{\Lambda}}{(1 + z)^2}\right],
\end{equation} with $\Omega_{\mathrm{k}} = 1 - \Omega_{\mathrm{m}} - \Omega_{\Lambda}$.
We then follow procedure given by \citep{csp01, Bi_1993} to generate density and velocity fields along line of sight using eqs.~\ref{eq:Pb_1d} and ~\ref{eq:Pv_1d}.

The lognormal approximation then accounts for the quasi-linear densities by assuming the baryonic density to be a lognormal variable,
\begin{equation}
     n_{\mathrm{b}}(x,z) = A~\mathrm{e}^{\delta^L_{\mathrm{b}}(x,z)}, \label{eq:LNapprox}
\end{equation}
where $A$ is a normalization constant fixed by setting the average value of $n_{\mathrm{b}}(x,z)$ to the mean baryonic density $\bar{n}_\mathrm{b}(z)$ at that redshift.

From the line of sight density and velocity, one can compute the neutral hydrogen field assuming the IGM to be in photoionization equilibrium,

\begin{equation}
     \alpha_A[T(x,z)]~n_{\mathrm{p}}(x,z)~n_{\mathrm{e}}(x,z) = n_{\mathrm{HI}}(x,z)~\Gamma_{12}(z)/(10^{12}~\mathrm{s}),
 \end{equation}
 where $\alpha_A(T)$ is the case-A recombination coefficient at temperature $T$, $n_\mathrm{p}, n_\mathrm{e}$ are the number densities of protons and free electrons respectively and $\Gamma_{\mathrm{12}}$ is the hydrogen photoionization rate (in units of $10^{-12}$ s$^{-1}$ and assumed to be homogeneous). Assuming a fully ionized IGM, $n_\mathrm{p}, n_\mathrm{e}$ are given by,

\begin{equation}
     n_p(x,z) = \frac{4(1 - Y)}{4 - 3Y}n_{\textrm{b}}(x,z)\, ; \, n_e = \frac{4 - 2Y}{4 - 3Y}n_{\textrm{b}}(x,z)
\end{equation}where $Y (\sim 0.24)$ is helium mass fraction.
We compute the line of sight IGM temperature using a power-law temperature-density relation characterized by the IGM temperature at mean density $T_0$ and the equation of state index $\gamma$, for the equation $T(x,z) = T_0(z)[1 + \delta^L_{\mathrm{b}}(x,z)]^{\gamma (z) - 1}$. The Ly$\alpha$ optical depth is then calculated by accounting for thermal and natural broadening at each grid point $x_i$,
\begin{align}
    \tau(x_i, z) &= \frac{c I_{\alpha}}{\sqrt{\pi}} \sum_j \delta x \frac{n_{\mathrm{HI}}(x_j,z)}{b(x_j,z)[1+z(x_j)]}
    \notag \\
    &\times  V_{\alpha}\left(\frac{c[z(x_j)-z(x_i)]}{b(x_j,z)[1+z(x_i)]}+\frac{v^L_{\mathrm{b}}(x_j,z)}{b(x_j,z)}\right),
\end{align}
where $\delta x$ is the the grid size, $I_{\alpha} = 4.45\, \times 10^{-18}$ cm$^2$ is the Ly$\alpha$ absorption cross section and $V_{\alpha}(\Delta v / b)$ is the Voigt profile for the Ly$\alpha$ transition and 
\begin{equation}
    b(x,z) = \sqrt{\frac{2 k_{\mathrm{boltz}} T(x,z)}{m_{\mathrm{p}}}},
\end{equation}
where $m_\mathrm{p}$ is the proton mass. We can then calculate the main observable, i.e., the normalized Ly$\alpha$ transmitted flux, $F(x_i, z) = e^{-\tau(x_i, z)}$. To mimic observational data, we also convolve $F(x_i, z)$ with Gaussian line spread function of full width at half maximum 7 km s$^{-1}$ as well as add random noise of SNR 50 per pixel. \textit{Our model is thus described by four free parameters, namely, \{$x_\mathrm{J}, T_0, \gamma, \Gamma_{12}$\}}. 

We now introduce a new parameter, $\delta z$, such that
\begin{equation}
z_{\mathrm{model}} = z_{\mathrm{SPH}} - \delta z
\end{equation}
where $z_{\mathrm{SPH}}$ is the redshift of the SPH data and $z_{\mathrm{model}}$ is the redshift at which lognormal model is evolved. Usually, in \citetalias{Arya_2023, Arya_2024}, we take $z_{\mathrm{model}} = z_{\mathrm{SPH}}$, which we have seen leads to over-prediction of Ly$\alpha$ absorbers (and therefore, $\Gamma_{12}$) compared to SPH. To mitigate this, we simulate the lognormal spectra at a lower redshift. This allows for reduction in densities that give rise to Ly$\alpha$ forest in lognormal model, with densities shifting to under- and over-dense regions, and rectifies the poor recovery of $\Gamma_{12}$ between $2.2 \leq z \leq 2.7$ as seen in \citetalias{Arya_2024} to a significant extent. For this work, we fix $\delta z = 0.6$ the reason of which we discuss in \cref{sec:result}.

\subsection{SPH simulation}
We use publicly available Sherwood simulations suite \citep{bolton+17-sherwood} to test the validity of our model. These simulations were performed with a modified version of the cosmological smoothed particle hydrodynamics code P-Gadget-3, an extended version of publicly available GADGET-2 code \citep{Springel_2005}\footnote{\url{https://wwwmpa.mpa-garching.mpg.de/gadget/}}. The Sherwood suite is a collection of cosmological simulation boxes with volume
ranging from $10^3$ to $160^3$ $h^{-3}\,\textrm{cMpc}^3$ and number of particles ranging from $2\times 512^3$ to $2 \times 2048^3$. The size and resolution of simulation box are suitable for studying the small scale structures probed by Ly$\alpha$ forest.
The properties of Ly$\alpha$ forest from Sherwood simulation suite are well converged \citep{bolton+17-sherwood}. 
Similar to \citetalias{Arya_2024}, as the default, we choose a box of volume $40^3$ $h^{-3}\,\textrm{cMpc}^3$ containing $2048^3$ dark matter and baryonic particles each.

\subsection{Skewer configuration and covariance matrices}
The calculation of all relevant statistics including the mean flux, FPS and their covariances for both SPH and lognormal remains identical to \citetalias{Arya_2024} (see their section~2.3). We briefly describe the procedure here and request the readers to refer to \citetalias{Arya_2024} for more details. We calculate our "data points" by averaging statistics over 100 skewers picked randomly from a total 5000 available. The SPH covariance matrix is calculated using Jackknife resampling using the entire 50 (=5000/100) realizations. We artificially scale the errors on mean flux at every redshift to 5\% of mean flux since observed mean flux typically has $\sim 5\%$ error arising due to systematic uncertainty in continuum placement \citep{Gaikwad_2021}. For lognormal covariance matrix, we generate 40000 skewers of same size of that of SPH. To reduce uncertainty from lognormal relative to SPH, the covariance matrix for lognormal is calculated by averaging statistics over 200 skewers. The covariance matrix is then calculated using 200 (=40000/200) realizations without Jackknife resampling. For likelihood analysis, we use mean flux and FPS as statistics.

We have run six  Markov Chain Monte Carlo (MCMC)  chains using publicly available code \texttt{cobaya}\footnote{\url{https://cobaya.readthedocs.io/en/latest/sampler_mcmc.html}}\citep{Lewis_2002, Lewis_2013, Torrado_2021}, at $z =$ \{2.2, 2.3, 2.4, 2.5, 2.6, 2.7\}. To determine when a chain is converged, we use Gelman-Rubin statistics parameter, $R-1 < 0.05$ \citep{gelman_rubin, Lewis_2013}. The convergence for each chain takes $\sim 3$ days, using 64 cpus on a 2048 grid. 
All MCMC calculations were performed on the PARAM Sanganak cluster at IIT Kanpur.\footnote{\url{https://www.iitk.ac.in/new/param-sanganak}}

\section{Results}
\label{sec:result}

In this section, we present the recovery of the free parameters of the lognormal by comparing with the SPH simulations alongwith a brief discussion on improvement in recovery of $\Gamma_{12}$. 

\begin{figure*}
\centering
\includegraphics[width=\textwidth]{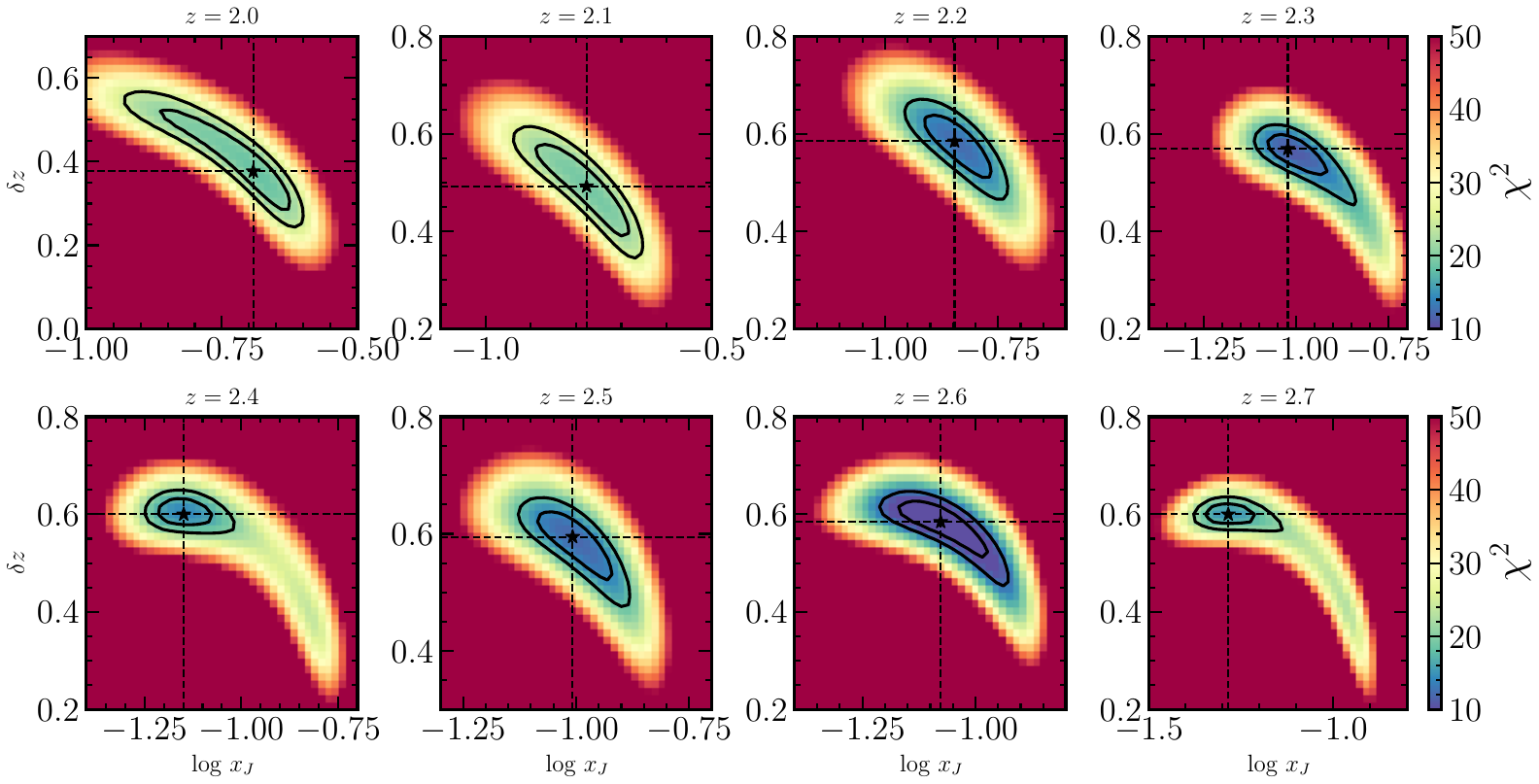}
\caption{$\chi^2$ colormap on log $x_{\textrm{J}}$ - $\delta z$ grid with \{$T_0, \gamma, \Gamma_{12}$\} fixed to their true values for all 8 redshift bins. We get acceptable fits for $z \geq 2.2$. Black contours show 1 and 2-$\sigma$ confidence levels and black stars show position of best-fit \{$x_{\mathrm{J}}$, $\delta z$\}.}
\label{fig:2d_chi2}
\end{figure*}

\begin{figure*}
\centering
\includegraphics[width=\textwidth]{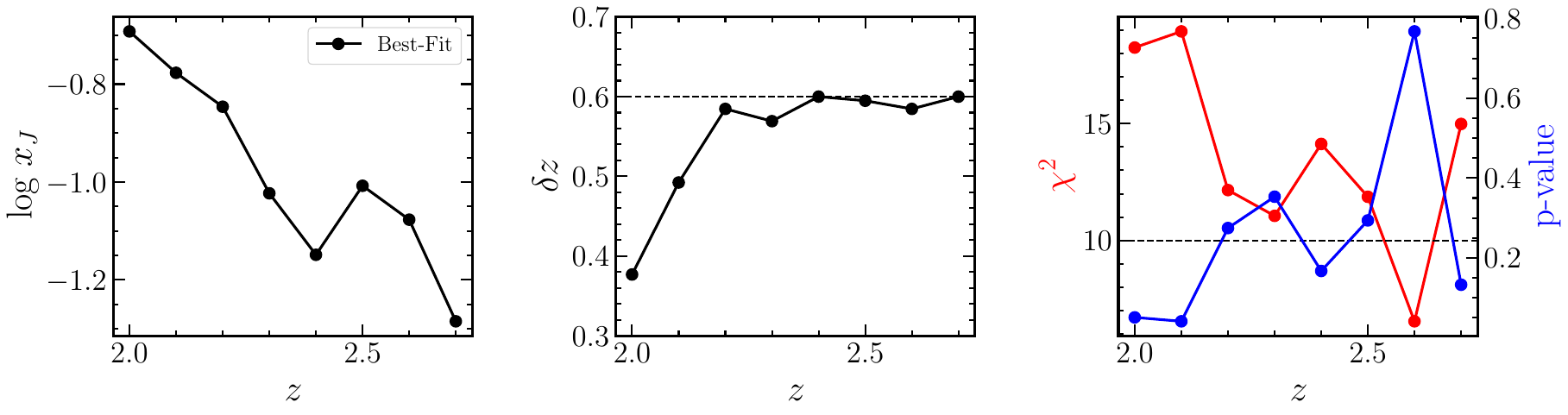}
\caption{First and second panels show redshift evolution of best-fit $x_{\textrm{J}}$ and $\delta z$ from 2D $\chi^2$ analysis and true values of \{$T_0, \gamma, \Gamma_{12}$\}. Third panel shows minimum $\chi^2$ (red) and corresponding p-value (blue) at each redshift. Black dashed line in the third panel shows the number of degrees of freedom.}
\label{fig:2d_chi2_evol}
\end{figure*}

\subsection{2-parameter fit}

Before presenting full-fledged MCMC results for our 4-parameter model described in the previous sections, we first try to estimate what would be the typical value of the Jeans length $x_{\mathrm{J}}$ and $\delta z$, parameters which do not have an obvious counterpart in the SPH simulations. To this end, we do a simple $\chi^2$-minimization using a 2D grid in $\log$ $x_{\mathrm{J}}$ - $\delta z$ and find the values of \{$x_{\mathrm{J}}$, $\delta z$\} which best fit the simulation statistics. For the other three parameters, $\Gamma_{12}, T_0, \gamma$, we use their corresponding true values in the SPH simulation. As mentioned earlier, we use the two flux statistics $\bar F$ and FPS for calculating likelihood analysis. We repeat this exercise at all the 8 redshifts.

Fig.~\ref{fig:2d_chi2} shows the colormap plot of $\chi^2$ as a function of $x_{\mathrm{J}}$ and $\delta z$ for all 8 redshifts, $z =$ \{2, 2.1, 2.2, 2.3, 2.4, 2.5, 2.6, 2.7\}.
Fig.~\ref{fig:2d_chi2_evol} summarises the redshift evolution of best-fit $x_{\mathrm{J}}$ and $\delta z$. At every redshift, we see clear minima in Fig. \ref{fig:2d_chi2} which provides overall decreasing (\textit{w.r.t.} redshift) best-fit values of $x_{\textrm{J}}$ from $\sim$ 200 $\kpch$ at $z = 2.0$ to $\sim$ 50 $\kpch$ at $z = 2.7$ in Fig.~\ref{fig:2d_chi2_evol}. This value is slightly smaller than the values obtained in \citetalias{Arya_2024} but of the same order as the ones obtained by assuming Ly$\alpha$ absorbers to be in hydrostatic equilibrium at a temperature $\sim 10^4$K \citep{schaye_2001} and from Bayesian formalism based on phase angle PDF using close quasar pair dataset \citep{rorai_2013}. The best-fit values of $\delta z$, however, show a different trend. For the first two redshift bins, values of $\delta z$ are $\sim 0.4$ and $0.5$ respectively but for $z \geq 2.2$, $\delta z$ remains roughly constant, $\sim 0.6$. Additionally, the minimum reduced $\chi^2$, $\chi^2_{\nu,\textrm{min}}$ and p-values for $2 \leq z \leq 2.7$ are \{1.8, 1.9, 1.2, 1.1, 1.4, 1.2, 0.6, 1.5\} and \{0.05, 0.05, 0.3, 0.35, 0.15, 0.8, 0.15\} respectively, implying that the fits are reasonably acceptable for $z \geq 2.2$. The reason behind flattening of $\delta z$ for $z \geq 2.2$ is still unclear, but we use it to our advantage by fixing $\delta z = 0.6$ at these redshifts instead of treating it like a free parameter in MCMC runs. This not only simplifies our model by removing any possible degeneracies arising due to $\delta z$ (similar to $\nu$ and $\Gamma_{12}$ in \citetalias{Arya_2024}) but also speeds up the computing time by a factor $\sim 4$. We also argue that the modified lognormal described in \citetalias{Arya_2024} produced good fits for the data and recovered all the three parameters within $\lesssim 1-\sigma$ for $z \epsilon \{2, 2.1\}$ with the over-prediction of $\Gamma_{12}$ occurring only at $z > 2.1$.

Keeping this in mind, we proceed to vary all the model parameters simultaneously in the next section. We will return to a discussion of the quality of parameter recovery in \cref{subsec:recovery}.

\subsection{4-parameter fit}

\begin{figure*}
\centering
\includegraphics[width=\textwidth]{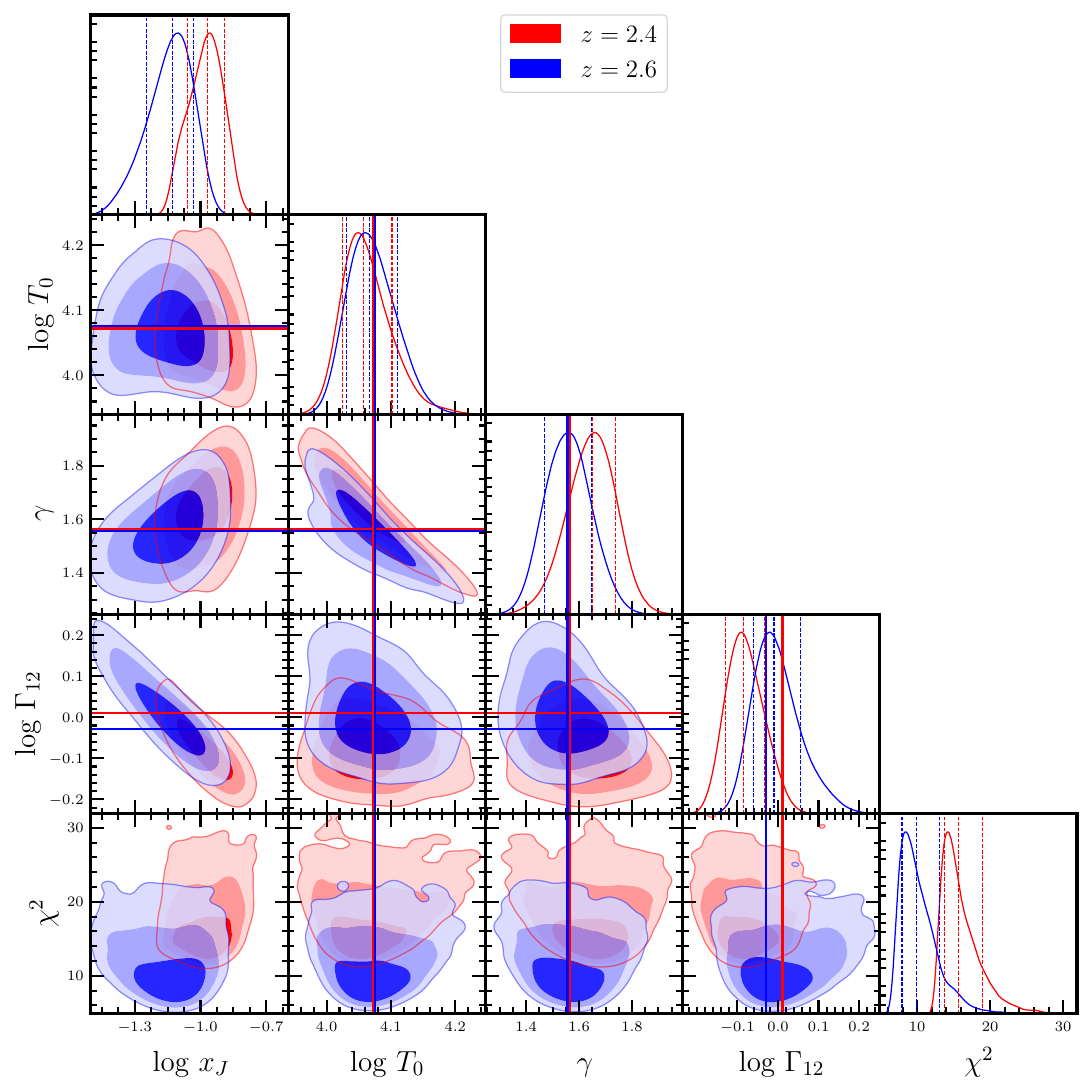}
\caption{1, 2, and 3$-\sigma$ contours for redshifts which produce best ($z = 2.6$) and worst ($z = 2.4$) fits. Colour coded horizontal and vertical solid lines show true values of parameters and dashed lines show 16, 50,and 84 percentiles.}
\label{fig:corner_main}
\end{figure*}

\begin{figure*}
\centering
\includegraphics[width=\textwidth]{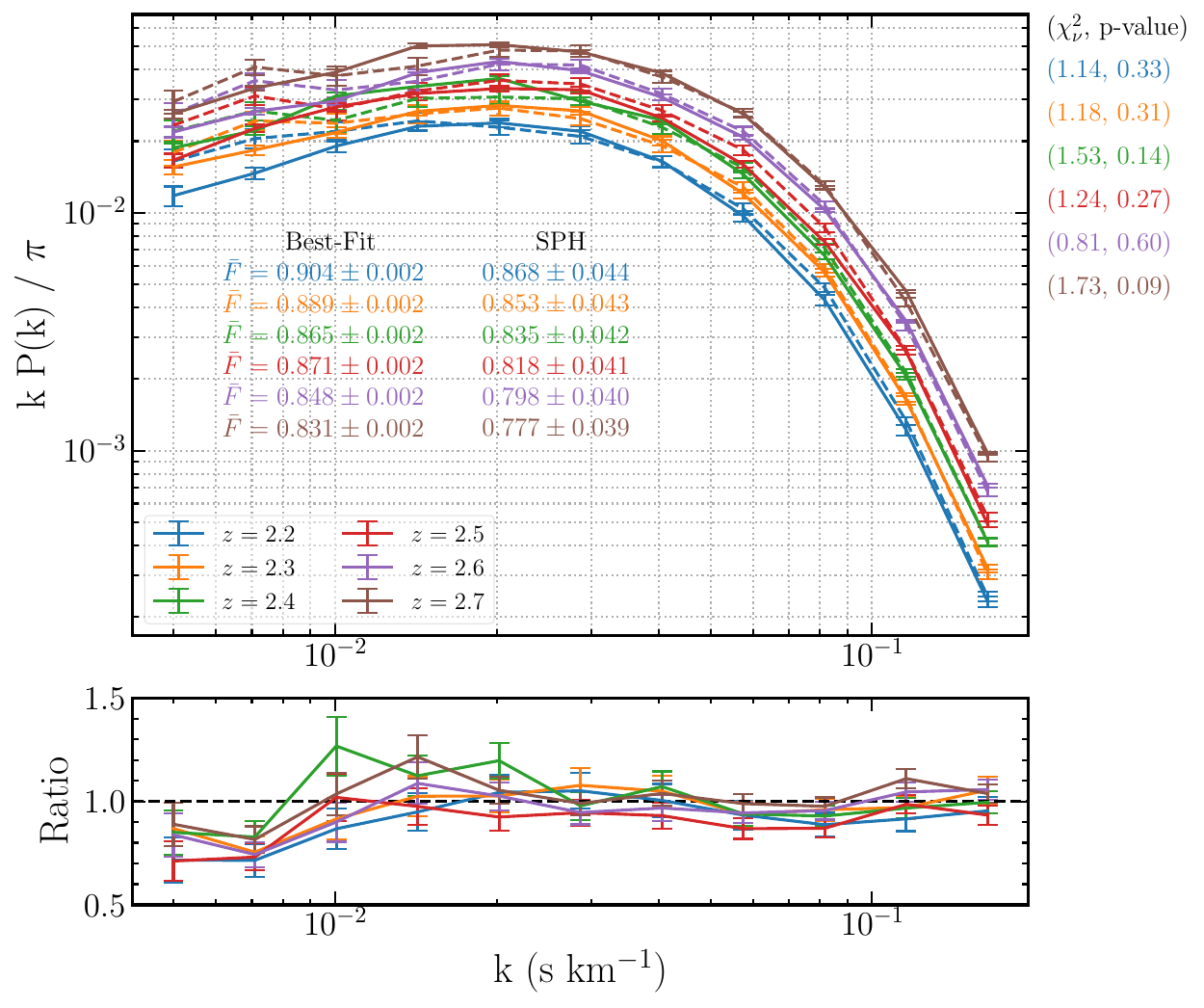}
\caption{Flux statistics for best-fit model and SPH data for all the redshifts.}
\label{fig:stat_main}
\end{figure*}

\begin{figure*}
\centering
\includegraphics[width=\textwidth]{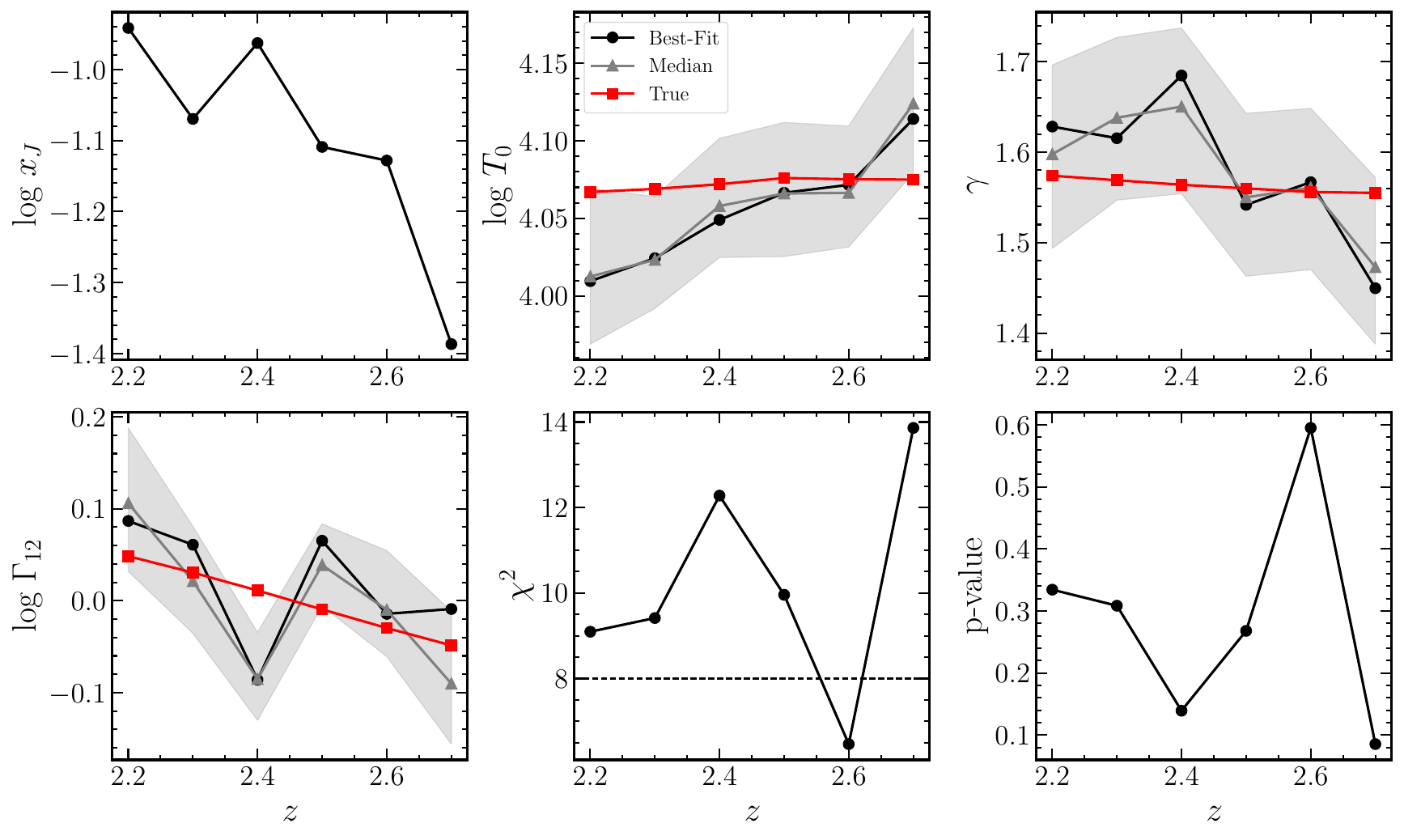}
\caption{Redshift evolution of parameters and $\chi^2$ shown with black circles. Gray triangles and shaded regions are median and (16, 84) percentiles from MCMC chains respectively. Red squares show true values of parameters in SPH.}
\label{fig:param_evol}
\end{figure*}

In this section, we describe the results from MCMC runs where all the free parameters are allowed to vary. We also discuss the improvements in recovery of parameters compared to \citetalias{Arya_2024}. Table \ref{table:priors} lists the priors on parameters, $\{\textrm{log}\, x_{\textrm{J}},\, \textrm{log}\, T_0, \, \gamma,\, \textrm{log}\, \Gamma_{12}\}$. Similar to \citetalias{Arya_2024}, we have used simple and flat, albeit narrower priors on the three parameters, $\{\textrm{log}\, T_0, \, \gamma,\, \textrm{log}\, \Gamma_{12}\}$. On log $x_{\textrm{J}}$, however, we use a more physically motivated prior by calculating the lower limit on the prior, $x_{\textrm{J,th}}$, using equation
\begin{equation}
    x_{\textrm{J,th}} = \frac{1}{H_0}\left[\frac{2\gamma_{\textrm{s}} k_{\textrm{B}}T_{0,\textrm{s}}}{3\mu m_{\textrm{p}}\Omega_{\textrm{m}}(1+z)}\right]^{1/2}
    \label{eq:xJ_th}
\end{equation}
where $T_{0,s}$ and $\gamma_{s}$ are values of $T_0$ and $\gamma$ sampled in the MCMC chain respectively. For reference, the values of log $x_{\textrm{J,th}}$ at redshifts \{2.2, 2.3, 2.4, 2.5, 2.6, 2.7\} for corresponding true values of $T_0$ and $\gamma$ are \{-0.879, -0.883, -0.895, -0.900, -0.905\} respectively. The choice for narrowing the priors is just to speed up the computation and does not affect the quality of results.
The motivation behind imposing the limit from eq.~\ref{eq:xJ_th} was to disallow lognormal model from converging to unphysically small values of $x_{\textrm{J}}$ ($\sim 10\, \kpch$) at high redshifts. However, as discussed earlier, in the present setup, the best-fit values of $x_{\textrm{J}}$ tend to be smaller than the ones obtained in \citetalias{Arya_2024}. To allow the lognormal model to access these smaller values in $x_{\textrm{J}}$ while keeping the chains within physical range, we relax the lower limit to 0.25$x_{\textrm{J,th}}$.

\begin{table}
\centering
\begin{tabular}{||c c||} 
\hline
Parameter & Prior \\ [0.5ex]
\hline\hline
log $x_{\textrm{J}}$ & [log $x_{\textrm{J,th}}$, -0.5]\\ 
\hline
log $T_0$ & [3, 5]\\ 
\hline
$\gamma$ & [0.5, 3]\\
\hline
log $\Gamma_{12}$ & [-1, 0.5]\\
\hline
\end{tabular}
\caption{Priors on parameters, $\{\textrm{log}\, x_{\textrm{J}},\,\textrm{log}\, T_0, \, \gamma,\, \textrm{log}\, \Gamma_{12},\, \nu\}$, where log $x_{\textrm{J,th}}$ is calculated using eq.~\ref{eq:xJ_th}. See text  for a discussion of the parameter $x_{\rm J}$.}
\label{table:priors}
\end{table}

The true values (i.e., the values used in or obtained from the SPH simulations) alongwith best-fit and median of the parameters at each redshift are reported in Table \ref{table:true_val_table}. In figs.~\ref{fig:corner_main}, \ref{fig:stat_main}, and \ref{fig:param_evol} we show the contour plots (68.3, 95.4, 99.7 percentiles) for two redshifts, $z = 2.4 \, \textrm{and} \, 2.6$ obtained from MCMC runs, corresponding best-fit and SPH flux statistics, and the evolution of best-fit values of parameters with redshift respectively. We choose these two redshifts as they produce the worst and best quality fits respectively. Fig.~\ref{fig:stat_main} shows that we get good fits at all redshifts, with minimum $\chi^2 < 15$ for 8 degrees of freedom. We do however, note that the model consistently underestimates power in the two smallest $k-$bins as well as overestimates mean transmitted flux at all redshifts. From fig.~\ref{fig:param_evol}, it is evident that the lognormal model does a good job at recovering all the three IGM parameters $T_0$, $\gamma$ and $\Gamma_{12}$ at all redshifts with the true values being within 1$-\sigma$ from the median. Compared to \citetalias{Arya_2024}, we find three major upswings regarding parameter recovery in this work. Firstly, the present work is able to recover the thermal history more accurately, with the errors decreasing by $\sim 20\%$ and $\sim 70\%$ on $T_0$ and $\gamma$ respectively. Secondly, \citetalias{Arya_2024} had presented a bimodal distribution of $\gamma$ for $z \geq 2.5$, with an "inverted" $T-\Delta_{\textrm{b}}$ relation at $z=2.7$. As evident, the trailing lognormal model is able to do away with both these issues. Lastly and most importantly, the best-fit (median) value of $\Gamma_{12}$ is at $\lesssim 10\%$ ($\sim 1-\sigma$) from the true values at all redshifts, except $z=2.4$, where it is undervalued at $15\%$ ($\sim 1.5-\sigma$). This is in contrast to \citetalias{Arya_2024} where $\Gamma_{12}$ was being recovered at $\sim$ 50-80\% ($> 3-\sigma$) for $z \geq 2.2$. We also see that the shape of the curve of redshift evolution of $\Gamma_{12}$ now follows closer to that of SPH.

Another difference lies in the change in the degeneracy structure between $x_{\textrm{J}}$ and $\Gamma_{12}$. compared to \citetalias{Arya_2024} which displayed a redshift-depended relation where there is a strong positive degeneracy at $z = 2.7$ but gradually turns mildly negative with decreasing redshift. The present work shows consistently negative degeneracy at all redshifts (see fig.~\ref{fig:corner_main}). Such degeneracy structure is reasonable since increasing (decreasing) $x_{\textrm{J}}$ will erase (enhance) baryonic density fluctuations at small scales. Thus, the model tries to counteract the effect by producing more (less) neutral hydrogen by decreasing (increasing) $\Gamma_{12}$.

\begin{table*}
\centering
\scalebox{0.7}{
\begin{tabular}{||c c c c c||} 
\hline
Redshift & log $x_{\textrm{J}}$ ($\Mpch$) & log $T_0$ (K) & $\gamma$ & log $\Gamma_{12}$ (10$^{-12}$ s$^{-1}$) \\[0.8ex]
\hline\hline
2.2 & - / -0.94(-0.97$^{+0.12}_{-0.12}$) & 4.07 / 4.01(4.01$^{+0.06}_{-0.04}$) & 1.57 / 1.63(1.60$^{+0.10}_{-0.11}$) & 0.05 / 0.09(0.11$^{+0.08}_{-0.07}$) \\[0.8ex]
\hline
2.3 & - / -1.07(-1.01$^{+0.10}_{-0.09}$) & 4.07 / 4.02(4.02$^{+0.04}_{-0.03}$) & 1.57 / 1.62(1.64$^{+0.09}_{-0.09}$) & 0.03 / 0.06(0.02$^{+0.06}_{-0.06}$) \\[0.8ex]
\hline
2.4 & - / -0.96(-0.97$^{+0.08}_{-0.09}$) & 4.07 / 4.05(4.06$^{+0.04}_{-0.03}$) & 1.56 / 1.68(1.65$^{+0.09}_{-0.10}$) & 0.01 / -0.09(-0.09$^{+0.05}_{-0.04}$) \\[0.8ex]
\hline
2.5 & - / -1.11(-1.06$^{+0.07}_{-0.06}$) & 4.08 / 4.07(4.06$^{+0.05}_{-0.05}$) & 1.56 / 1.54(1.55$^{+0.09}_{-0.09}$) & -0.01 / 0.07(0.04$^{+0.04}_{-0.04}$) \\[0.8ex]
\hline
2.6 & - / -1.13(-1.12$^{+0.09}_{-0.13}$) & 4.08 / 4.07(4.07$^{+0.04}_{-0.054}$) & 1.56 / 1.57(1.56$^{+0.09}_{-0.09}$) & -0.03 / -0.01(-0.01$^{+0.07}_{-0.05}$) \\[0.8ex]
\hline
2.7 & - / -1.39(-1.25$^{+0.13}_{-0.13}$) & 4.08 / 4.11(4.12$^{+0.05}_{-0.04}$) & 1.56 / 1.45(1.47$^{+0.10}_{-0.09}$) & -0.05 / -0.01(-0.09$^{+0.08}_{-0.07}$) \\[0.8ex]
\hline
\end{tabular}}
\caption{True / best-fit(median) values for the parameters explored in MCMC run. Please see that $x_{\textrm{J}}$ does not have any "true" value.}
\label{table:true_val_table}
\end{table*}

\subsection{Recovery of $\Gamma_{12}$}
\label{subsec:recovery}

\begin{figure*}
\centering
\includegraphics[width=\textwidth]{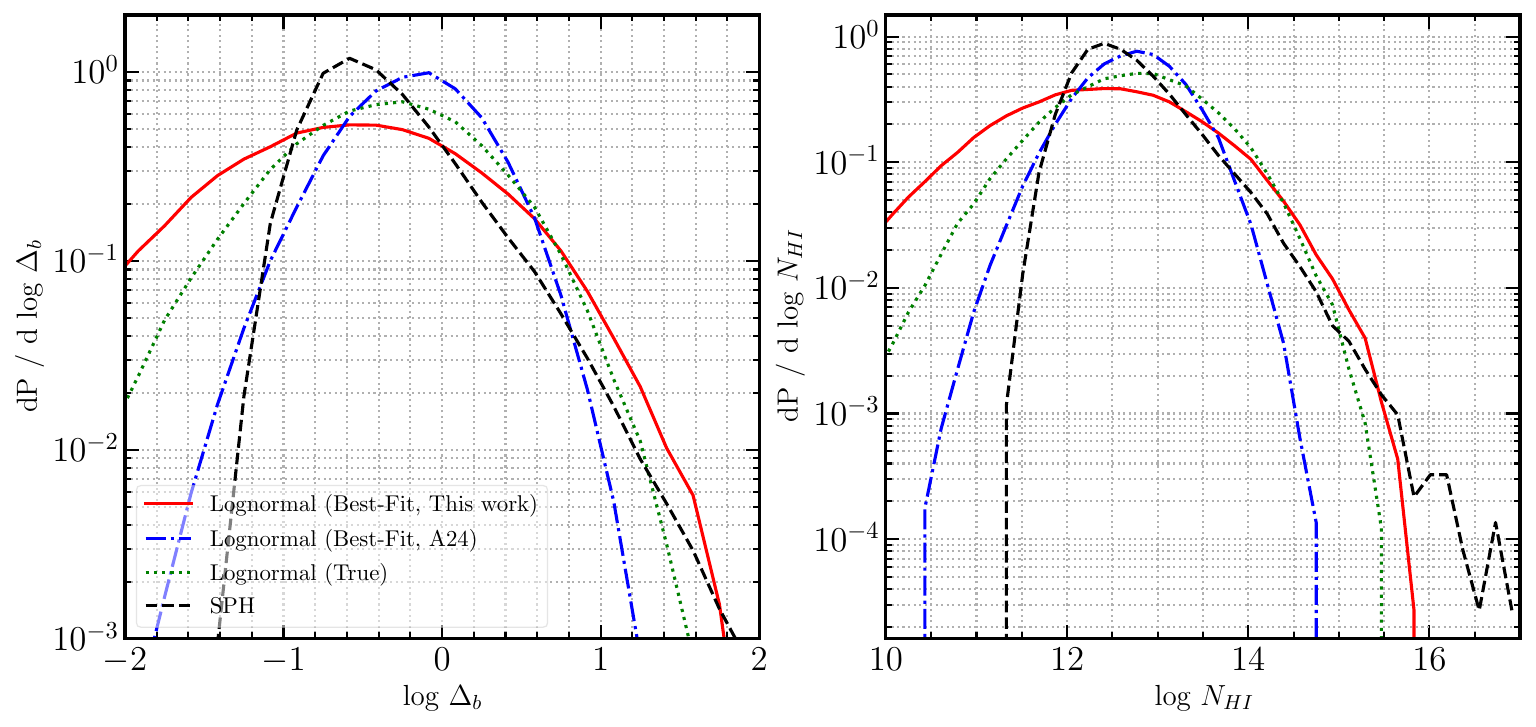}
\caption{Comparison of $\Delta_{\textrm{b}}$ (left) and $N_{\textrm{HI}}$ (right) in SPH to lognormal at $z=2.5$ with different parameter choices, (i) best-fit parameters from this work (red solid), (ii) best-fit parameters from \citetalias{Arya_2024} (blue dash-dot), and (iii) true parameters (green dotted).}
\label{fig:Deltab_NHI}
\end{figure*}

In here, we present a brief qualitative assessment about the improvement in the recovery of $\Gamma_{12}$. Fig.~\ref{fig:Deltab_NHI} shows comparisons of PDFs of 1D baryonic density fields $\Delta_{\textrm{b}}$ (left panel) and column density $N_{\textrm{HI}}$ (right panel) at $z=2.5$ between SPH (black dashed) and three models of lognormal approximation, (i) best-fit parameters from this work (red solid), (ii) best-fit parameters from \citetalias{Arya_2024} (blue dash-dot) and (iii) lognormal with true parameters (green dotted). We use the analytic expression from \citep{schaye_2001} to calculate column densities,
\begin{equation}
    N_{\textrm{HI}} \sim 2.7 \times 10^{13} \textrm{cm}^{-2} \Delta_{\textrm{b}}^{3/2} T_4^{-0.26} \Gamma_{12}^{-1} \left(\frac{1+z}{4}\right)^{9/2} \left(\frac{\Omega_{\textrm{b}}h^2}{0.02}\right)^{3/2} \left(\frac{f_\textrm{g}}{0.16}\right)^{1/2}
\end{equation}
Most of Ly$\alpha$ forest typically arises from mildly non-linear densities, $\Delta_{\textrm{b}} \sim 1-10$ $(N_{\textrm{HI}} \sim 10^{13}-10^{14.7} \textrm{cm}^{-2})$. From fig.~\ref{fig:Deltab_NHI}, it is evident that the lognormal model (run with true parameters) over-produces such regions compared to SPH. Hence, the model within an MCMC setup overestimates the photoionization rate to reduce the number of excess absorbers, as seen in \citetalias{Arya_2024}. An alternative way to approach this issue is by simulating the Ly$\alpha$ spectra in the lognormal model at a lower redshift, which leads to lower Ly$\alpha$ absorbers even with similar $\Gamma_{12}$. This is equivalent to evolving the lognormal density fields for a longer time than the simulations. As evident in fig.~\ref{fig:Deltab_NHI}, such an exercise leads to flattening of both the PDFs. While both the under-dense and dense regions are overestimated compared to \citetalias{Arya_2024}, the amount of mild over-densities primarily responsible for the forest are reduced. We speculate that this is related to the fact that lognormal model is unable to properly account for non-linearities in structure formation, due to the log-density field being Gaussian and therefore ignoring the higher order moments.

\section{Conclusions}
\label{sec:conclude}

The advent of high quality QSO spectra from large cosmological surveys necessitate fast and accurate semi-analytical modelling of the IGM. The objective of this work is to build upon \citetalias{Arya_2024} and alleviate the poor recovery of ionization history, $\Gamma_{12}$ of the IGM for $2.2 \leq z \leq 2.7$ in the lognormal model of the baryonic density. Similar to, \citetalias{Arya_2024}, we employ our end-to-end MCMC technique to carry out the parameter estimation using two transmitted flux statistics: the mean flux and the flux power spectrum.

We find that the modified lognormal model where the 1D Gaussian baryonic density is scaled by parameter $\nu$, $\delta^L_{\mathrm{b}} \rightarrow \nu \delta^L_{\mathrm{b}}$ can recover the thermal parameters, $T_0$ and $\gamma$ reliably but overestimates $\Gamma_{12}$ by $\gtrsim 3-\sigma$ for $z \geq 2.2$. This is owing to the fact that the model is a poor description of the underlying baryonic density PDF obtained from the Sherwood simulations and over-produces the Ly$\alpha$ absorbers. We mitigate this drawback by simulating the Ly$\alpha$ spectra in the model at a lower redshift than the simulations, $z_{\mathrm{model}} = z_{\mathrm{SPH}} - \delta z$. This modification is a better reflection of SPH than the parameter $\nu$. In particular, $\Gamma_{12}$ is recovered within $1-\sigma$ for all redshift bins except $z=2.4$, where it is discrepant at $\sim 1.5-\sigma$. The recovery of thermal history is also significantly alleviated with the temperature-density relation no longer being "inverted" at higher redshift bins.

Another advantage of this model is that $\delta z$ plateaus at $\sim 0.6$ within the redshift range, allowing us to treat it like a fixed parameter. This not only simplifies the model by avoiding any possible degeneracies with other IGM parameters but also speeds up the convergence of MCMC runs.
We also explore the reason for the improvement in parameter recovery in some detail, and speculate that the absence of higher order moments in log-baryonic density field in the model causes it to over-produce mildly non-linear densities ($\Delta_{\mathrm{b}} \sim 1-10$) which are the primary source of Ly$\alpha$ forest. Therefore, evolving the model at a lower redshift allows it to reduce the excess Ly$\alpha$ absorbers without overestimating $\Gamma_{12}$.

This work opens up the possibility of using lognormal approximation to explore a joint astrophysical and cosmological parameter space as well as parameters of various dark matter models, more efficiently generate mock catalogues and calculate covariance matrices for large volume cosmological surveys such as DESI, WEAVE etc. Additionally, one can also use lognormal to set parameters for initial sampling in full hydrodynamical simulations, and/or narrow down the range of priors thus reducing significant computing time. It would also be interesting to explore the idea of treating $\delta z$ like a free parameter in MCMC runs which might allow to extend the usable redshift range of lognormal model. We leave such an exercise for the future.

\section*{Acknowledgments}

I sincerely thank Aseem Paranjape and Tirthankar Roy Choudhury for their useful discussions.
I gratefully acknowledge use of the Param Sanganak, the High Performance Computing (HPC) facility, IIT Kanpur. I also thank the Sherwood simulation team for making their data publicly available.

\section*{Data Availability}

The Sherwood simulations are publicly available at \url{https://www.nottingham.ac.uk/astronomy/sherwood/index.php}. The data generated during this work will be made available upon reasonable request to the authors.

\bibliographystyle{JHEP}
\bibliography{references}

\label{lastpage}

\end{document}